\documentclass[final,5p,times,twocolumn]{elsarticle}

\usepackage{amssymb,color}

\journal{Physica C}

\begin{document}

\begin{frontmatter}



\title{Evidence for Pauli-limiting behaviour at high fields and enhanced upper
critical fields near $T_c$ in several 
disordered FeAs based Superconductors}


\author[1]{G.\ Fuchs}
\ead{fuchs@ifw-dresden.de}
\author[1]{S.-L.\ Drechsler}
\author[1]{N.\ Kozlova}
\author[2]{M.\ Bartkowiak}
\author[1]{G.\ Behr}
\author[1]{K.\ Nenkov}
\author[3]{H.-H.\ Klauss}
\author[1]{J.\ Freudenberger}
\author[1]{M.\ Knupfer}
\author[1]{F.\ Hammerath}
\author[1]{G.\ Lang}
\author[1]{H.-J.\ Grafe}
\author[1]{B.\ B\"uchner}
\author[1]{L.\ Schultz}
\address[1]{Leibniz-Institut, IFW-Dresden, 
P.O.\ Box 270116, D-01171 Dresden, Germany}
\address[2]{FZ Dresden-Rossendorf (FZD), Germany}
\address[3]{Institut f\"ur Festk\"orperphysik, 
TU Dresden, Germany}

\begin{abstract}
{\small
We report resistivity and upper critical  field $B_{c2}(T)$
 data for disordered (As deficient) 
 LaO$_{0.9}$F$_{0.1}$FeAs$_{1-\delta}$  
in a wide temperature and high field range up to 60~T. These 
samples exhibit a slightly enhanced  superconducting transition
at $T_c = 28.5$~K 
and a 
significantly enlarged slope d$B_{c2}$/d$T = -5.4$~T/K near $T_c$ 
which contrasts with a flattening of $B_{c2}(T)$ starting near 23~K above 
30~T. The latter evidences
Pauli limiting behaviour (PLB) with 
$B_{c2}(0)\approx 63$~T. We compare our results with $B_{c2}(T)$-data 
from
the literature for clean
 and disordered samples. Whereas clean samples show almost no PLB 
for fields below 60 to 70~T, the hitherto unexplained 
pronounced 
flattening of 
$B_{c2}(T)$
 for applied fields $H \parallel ab$ observed for several disordered closely 
related systems is interpreted also as a manifestation of PLB. 
Consequences 
are discussed in terms of disorder effects within 
the frames
 of (un)conventional 
superconductivity, respectively.}
\end{abstract}


\end{frontmatter}


FeAs based superconductors 
show
high
transition temperatures $T_c \leq 57$~K and remarkably
 high upper critical fields $B_{c2}(0)$ exceeding often 70~T. Many 
 of their
 basic 
properties 
and the underlying pairing mechanism are still 
not well understood. A study of $B_{c2}(T)$,
 in particular, studies
on disordered FeAs superconductors are of large interest since 
for an unconventional pairing both $T_c$ and d$B_{c2}$/d$T$ at $T_c$ 
might be reduced
by  introducing 
disorder. Here,
$B_{c2}(T)$
of As-deficient (AD)
LaO$_{0.9}$F$_{0.1}$FeAs$_{1-\delta}$ 
samples are
 studied in fields up to 60~T. 

Polycrystalline samples of LaO$_{0.9}$F$_{0.1}$FeAs were prepared by the 
standard solid state reaction 
\cite{Fuchs1}. Some samples have 
been wrapped in 
a Ta foil during the final annealing procedure. Ta acts as an As getter at 
high $T$ forming a solid solution of about 9.5 at.\% As in Ta. This 
leads to an As loss in the samples resulting in an As/Fe ratio $\sim$ 0.9. 
Due to disorder in the FeAs layer, 
a three times larger
resistivity above $T_c$ 
at 31~K is found for the 
studied AD samples
compared with
a clean reference sample.
Anyhow, 
the AD samples 
have, with $T_c = 28.5$~K, a {\it higher }
$T_c$ than that of stoichiometric reference 
samples (27.7~K) \cite{Fuchs1}. 
\begin{figure}[h]
\begin{center}
\includegraphics[width=0.8\linewidth, keepaspectratio]{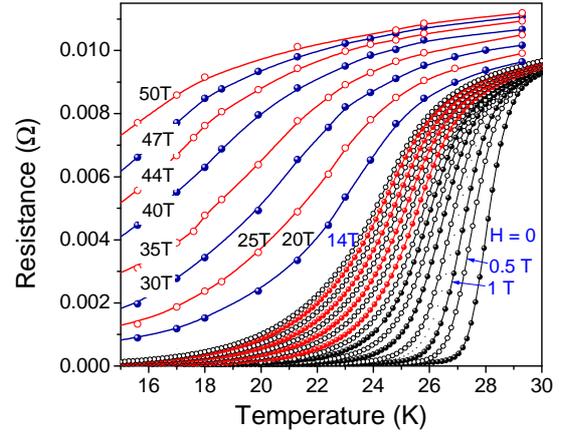}
\end{center}
\vspace{-0.5cm}
\caption{(Color online) $T$-dependence of  
the resistance from both DC and pulsed field measurements.}
\label{resistance}
\end{figure}
	
	For polycrystalline samples, only 
$B_{c2}^{ab}$ is accessible which is related to 
those grains oriented with their 
$ab$-planes along the applied field. $B_{c2}^{ab}$ was determined 
from 
resistance data 
by 
defining 
the onset 
of superconductivity (SC)  
at 90\% of the resistance $R_N$ in the normal 
state. The $T$-dependence of the resistance of a typical 
AD samples is shown in Fig.\ 1
for applied magnetic fields up to 50~T.
We confirmed that the pulsed field data
were unaffected by sample heating \cite{Fuchs2}.
$B_{c2}^{ab}(T)$ of our AD
sample obtained from pulsed field measurements in the IFW and the 
FZD is shown in Fig.\ 2 together with $B_{c2}$ data
reported for a clean reference sample \cite{Hunte}. The large slope 
d$B_{c2}/$d$T$= -5.4 T/K at $T_c$ of our AD samples
 points to strong impurity 
scattering in accord with enhanced resistivity at 30~K. For the clean 
sample \cite{Hunte} the available data up to 
45~T is well described by the WHH (Werthamer-Helfand-Hohenberg) 
model \cite{WHH}  
for the orbital limited $B^*_{c2}(T)$. Whereas for the AD
samples, the 
WHH model which predicts 
$B_{c2}^*(0) = 0.69T_c (\mbox{d}B_{c2}/\mbox{d}T)|_{T_c} = 106$~T at 
$T = 0$, fits the 
experimental data up to 30~T, only. For applied fields $>$
30~T increasing 
deviations from the WHH curve are clearly visible 
for the $B_{c2}(T)$
 data. 
The flattening of $B_{c2}(T)$ at high field 
points to its limitation by the Pauli spin paramagnetism. This effect is 
measured in the WHH model by the Maki parameter 
$\alpha = \sqrt{2} B_{c2}^*(0)/B_p(0)$, where 
$B_p(0)$ is the Pauli limiting field. The paramagnetically limited upper 
critical field, $B_{c2}^p$, is  given 
by $B_{c2}^p(0) = B_{c2}^*(0)(1+\alpha^2)^{-0.5}$. For our AD samples, a 
reasonable
fit of the experimental data to this model 
has been 
obtained 
for $\alpha = 1.31$ (see Fig.\ 2) 
and yields
$B_{c2}^p(0) = 63$~T. 
In contrast, $B_{c2}(T)$ data for clean LaO$_{0.93}$F$_{0.07}$FeAs 
samples \cite{Kohama} (see Fig.\ 3) show almost no PLB below 70 T.
\begin{figure}
\begin{center}
\includegraphics[width=0.92\linewidth, keepaspectratio]{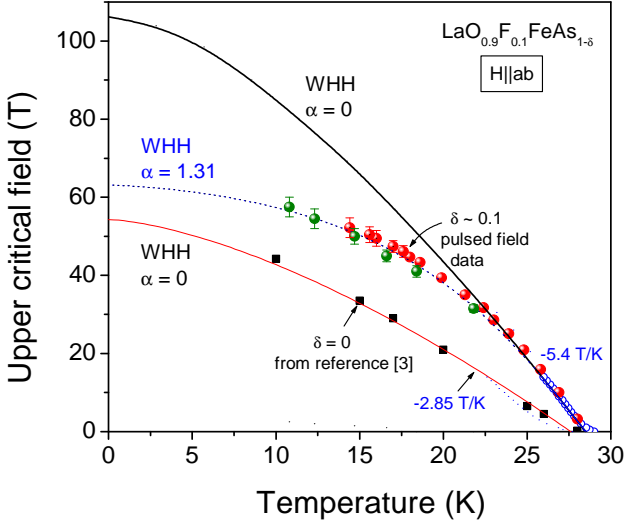}
\end{center}
\vspace{-0.5cm}
\caption{(Color online) $T$-dependence of  $B_{c2}^{ab}$. Data for the 
AD sample from 
DC ({\tiny $\blacksquare$}) and 
pulsed field measurements ( \textcolor{red}{$\bullet$}    - IFW Dresden,  
\textcolor{green}{$\bullet$}  - FZD) and data for a 
clean reference sample \cite{Hunte}. Solid lines: WHH model without PLB. 
Dotted line: $B_{c2}^*(T)$ for $\alpha = 1.31$ (see text).}
\label{compare}
\end{figure}
But for several possibly
disordered 
closely related systems, a similar flattening 
of $B_{c2}(T)$ as we found for our AD samples
 has been reported for applied 
fields $H \parallel ab$. This is shown in Fig.\ 3 
using normalized dimensionless units
$b^*(t) = B_{c2}(t)/\left[T_c(dB_{c2}/dT)|_{T_c}\right]$ 
and $t=T/T_c$.  
 The 
data in Fig.\ 3
are well described by the WHH model with the obtained Maki parameters 
$\alpha$. 
The deviation of $b^*(t)$ at low $T$ from $b^*(t)$  
for $\alpha = 0$ increases with $\alpha$
 due to rising paramagnetic pair-breaking.
In this context the partial substitution of Fe
by 4$d$ and 5$d$ transition metals (e.g.\ Ru, Rh, Ir, and Pd)
\cite{Schnelle}-
\cite{Ni}
is of interest. Due to their larger 
atomic
sizes compared
with that of Fe$^{2+}$ also a stronger disorder effect
compared with the 3$d$ substitutions by
Co and Ni should be expected. For all these cases we predict 
PLB, too, at variance with large WHH estimates 
on the basis of the enhanced initial slopes \cite{Qi,Han}.

	For AD samples we found 
	hints for an 
enhanced Pauli paramagnetism from $\mu$SR experiments 
\cite{Fuchs2}. Their improved SC at 
high $T$
and low fields could be understood within conventional $s_{++}$-wave SC 
by enhanced 
disorder. In contrast, for clean FeAs superconductors an unconventional 
$s_{\pm}$-wave 
scenario has been proposed \cite{Mazin08}. 
Based on our results for $B_{c2}(T)$, two 
alternative scenarios of opposite disorder effects might be suggested: 
(i) an impurity-driven change of the pairing state from $s_{\pm}$ to 
conventional $s_{++}$-wave SC and (ii) a 
special impurity-driven stabilization of the $s_{\pm}$ state where the 
As-vacancies are assumed to scatter mainly within the bands, only,
but not in between them. 
Concerning the impurity scattering and SC gap symmetry
a recent $^{75}$As-NMR study on a disordered AD sample 
\cite{Hammerath} revealed $T^{-1}_1\propto T^5$  for the spin-lattice 
relaxation rate
below $T_c$ compared with a $T^3$-law for a 'clean' sample
is of considerable interest.  

The PLB found here suggests to continue measurements at least 
up to 70~T 
to eludicate, whether there is still much room for 
increasing 
$B_{c2}$ beyond 80~T. Improving the low-field 
properties of FeAs superconductors by introducing As vacancies 
opens new preparation routes for optimising their properties.
 
{\it Note added.}
Finishing the present paper we have learned about a very interesting
$B_{c2}$-study by Lee {\it et al.} \cite{Lee09} 
on O deficient F-free and F doped
Sm-1111 single crystals with high $T_c$-values of 50.5 and 42~K, 
respectively.
Using the same approach and probing the same field range 
as we, the authors detected
for $B^{ab}_{c2}$ 
PLB, too 
($ \alpha = 2.3$ and 2.7). Thus, at present in total at least 
seven different
ferro-pnictide superconductors exhibit PLB
which now seems to become a rather general feature.

We thank I.\ Eremin and M.\ Korshunov for discussions.
\begin{figure}
\begin{center}
\includegraphics[width=0.92\linewidth, keepaspectratio]{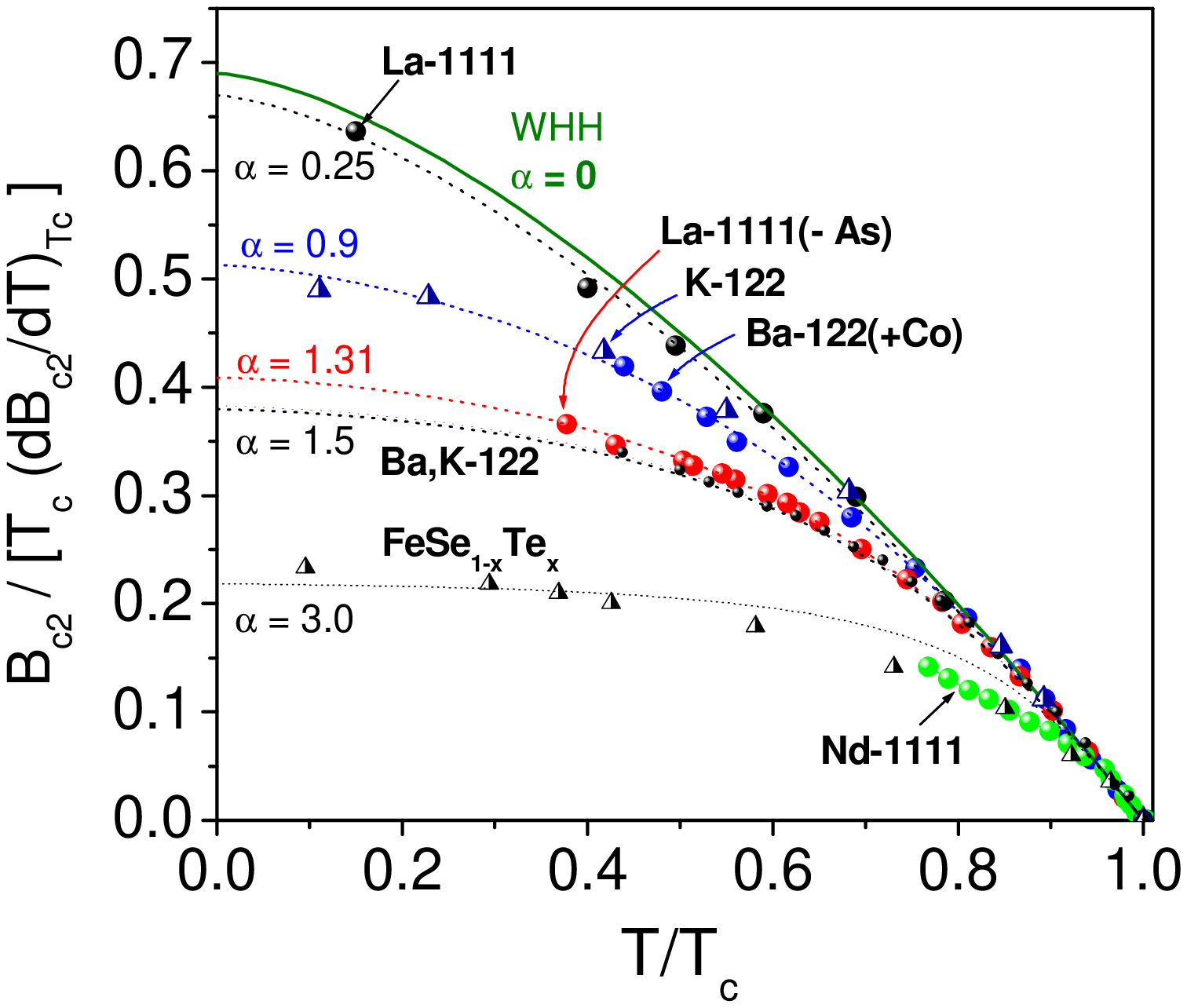}
\end{center}
\vspace{-0.5cm}
\caption{(Color online) Normalized upper critical field $b^*$
vs.\ $t=T/T_c$ 
for an AD LaO$_{0.9}$F$_{0.1}$FeAs$_{1-\delta}$ sample 
(La-1111(-As)) compared with data 
for non-deficient 
LaO$_{0.93}$F$_{0.07}$FeAs (La-1111, $T_c = 25$~K) \cite{Kohama}, 
Ba(Fe$_{0.9}$Co$_{0.1}$)$_2$As$_2$ (Ba-122(+Co), $T_c = 21.9$~K) 
\cite{Yamamoto}, KFe$_2$As$_2$ (K-122, $T_c = 2.8$~K) 
\cite{Terashima,remark}, 
Ba$_{0.55}$K$_{0.45}$Fe$_2$As$_2$ 
(Ba-122, $T_c = 32$~K) 
\cite{Altarawneh}, 
NdO$_{0.7}$F$_{0.3}$FeAs 
(Nd-1111, $T_c = 45.6$~K) 
\cite{Jaroszynski}, and
FeSe$_{0.25}$Te$_{0.75}$ $T_c = 14.2$~K) \cite{Kida,remark}. 
Dotted 
and solid lines: WHH model for the 
indicated  values. All curves for 
$H \parallel ab$.  }
\label{compare}
\end{figure}







\begin{thebibliography}{99}
\bibitem{remark}The original
$\alpha$-value 
as given by the authors differs
from the shown $\alpha$ obtained by us in 
re-analyzing their data according to our 
analysis \cite{Fuchs1,Fuchs2}.
\end{thebibliography}
\end{document}